\documentclass[final,3p,12pt,authoryear,a4paper]{elsarticle}

\usepackage{float}
\usepackage{setspace}
\usepackage[none]{hyphenat}
\usepackage{amsmath,amsfonts,amssymb,amsthm,pxfonts,dsfont}
\usepackage{natbib}
\usepackage[bookmarks, colorlinks]{hyperref}
\hypersetup{linkcolor=blue,citecolor=blue,filecolor=black,urlcolor=blue} 
\usepackage{graphicx} 
\usepackage{color}
\usepackage{geometry}
\usepackage{multirow}
\usepackage{booktabs}
\usepackage{rotating} 
\usepackage{subfig}
\usepackage[american]{babel}
\usepackage{longtable,lscape,ltcaption}
\usepackage{soul}
\usepackage[normalem]{ulem}
\usepackage{pdfpages}
\usepackage{orcidlink}

\emergencystretch=\maxdimen
\DeclareMathOperator{\diag}{diag}

\begin{document}

\biboptions{longnamesfirst}
\newtheorem{corollary}{Corollary}
\newdefinition{df}{Definition}
\newdefinition{algo}{Algorithm}

\journal{}

\begin{frontmatter}

\title{Identification Verification for\\ Structural Vector Autoregressions with Sparse\\ Heterogeneous Markov Switching Heteroskedasticity}

\author[gufs]{Fei Shang \textsuperscript{\orcidlink{0000-0003-1908-3275}}}
\author[um]{Tomasz Wo\'zniak\corref{cor1} \textsuperscript{\orcidlink{0000-0003-2212-2378}}}

\cortext[cor1]{Corresponding author. \emph{Email address:} \href{mailto:twozniak@unimelb.edu.au}{twozniak@unimelb.edu.au}.}

\address[gufs]{School of Economics and Trade, Guangdong University of Foreign Studies, Guangzhou 510006, China} 
\address[um]{University of Melbourne, 111 Barry St., 3053 Carlton, VIC, Australia}

\begin{abstract}
\noindent We propose a structural vector autoregressive model with a new and flexible specification of the volatility process which we call Sparse Heterogeneous Markov-Switching Heteroskedasticity. In this model, the conditional variance of each structural shock changes in time according to its own Markov process. Additionally, it features a sparse representation of Markov processes, in which the number of regimes is set to exceed that of the data-generating process, with some regimes allowed to have zero occurrences throughout the sample. We complement these developments with a definition of a new distribution for normalised conditional variances that facilitates Gibbs sampling and identification verification. In effect, our model: (i) normalises the system and estimates the structural parameters more precisely than popular alternatives; (ii) can be used to verify homoskedasticity reliably and, thus, inform identification through heteroskedasticity; and (iii) features excellent forecasting performance comparable with Stochastic Volatility. Finally, revisiting a~prominent macro-financial structural system, we provide evidence for the identification of the US monetary policy shock via heteroskedasticity, with estimates consistent with those reported in the literature.
\end{abstract}


\end{frontmatter}

\newpage
\section{Introduction}\label{sec:intro}

\noindent Heteroskedastic Structural Vector Autoregressions (SVARs) are commonly used because they can fit macro-financial aggregates and offer flexibility across a range of empirical applications. Some examples include influential studies such as that by \cite{sims2006}, which estimates an SVAR with Markov-switching heteroskedasticity for a~monetary system and interprets the occurrence of high and low volatility regimes in relation to the terms of particular Federal Reserve chairpersons. In other studies \citep[see e.g.][]{elder03,ccm18}, the estimated volatilities are interpreted as economic uncertainty and used to investigate the effects of uncertainty shocks. Heteroskedastic SVARs were used to identify all elements of the structural matrix and, therefore, to identify structural shocks through heteroskedasticity in the seminal studies by \cite{rigobon2003} and \cite{brunnermeier2021a}. Time-varying volatility has been documented to improve the forecasting performance of structural models \citep[see][and other papers cited in this introduction]{clark2015}, and to enable order-invariant representations that make forecasts even more precise \citep[see][]{chan2024}. Finally, this SVAR extension positively affects the model fit \cite[see e.g.][]{sims2006,lutkepohl2017,lutkepohl2018} and is documented to improve it by a greater margin than other extensions, such as time-varying parameters or hierarchical modelling \citep[see][]{ce18}.

The review of parametric specifications for structural shock heteroskedasticity in SVARs is opened by exogenous heteroskedastic regime changes in which the dates of volatility changes are fixed while the conditional variances are estimated as proposed by \cite{rigobon2003,brunnermeier2021a,lutkepohl2021}. Other more flexible regime change models include Markov-switching \cite[see e.g.][]{sims2006,lanne2010,lutkepohl2020} or Smooth Transition \citep[see][]{ln17} models in which the probabilities of regime occurrences are estimated as well. SVAR models in which conditional variances are modelled by a dynamic propagation equation include Stochastic Volatility (SV) \citep[see][]{ccm19,bb22,chan2024}, and Generalised Autoregressive Conditional Heteroskedasticity (GARCH) \citep[see e.g.][]{normandin2004,lm16} or their hybrids \citep[see e.g.][]{pajor2023}. Our listing is finalised by nonparametric specifications that imply heteroskedasticity via moment conditions \citep[see][]{ll21,g21,l21}. 

This extraordinary performance and multiplicity of modelling approaches come at various costs and challenges. SVARs with Stochastic Volatility that tend to outperform other specifications in terms of forecasting performance and model fit are computationally more demanding than, say, Markov-switching models. SV models are more difficult to estimate in the maximum likelihood than in the Bayesian approach, whereas the opposite is the case for the GARCH models. Regime change models suffer from a lack of identification due to label switching -- a likelihood-invariant permutation of regimes and regime-specific parameters. Structural models identified through heteroskedasticity are identified up to shock permutation. Finally, normalisation of the structural system is required because the structural matrix and the volatility process are multiplicative in the construction of the predictive density covariance matrix. It usually leads to parameter restrictions that make estimation more tedious.

In this context, our search for an adequate parametric specification is informed by the newest developments and encompasses most of the challenges. We propose a new volatility model for SVARs, which we call Heterogeneous Markov-Switching Heteroskedasticity (HMSH). We use the term \emph{heterogeneous} as each regime-specific parameter, that is, the structural shock's conditional variance, follows its own Markov process that determines its time-varying properties. It extends the existing approaches \cite[see][]{sims2006,lanne2010,lutkepohl2020} that impose the assumption of homogeneity, assuming that all conditional variances evolve according to one Markov process. By allowing for heterogeneity, our model is more flexible and aligns with the setup commonly adopted in SVARs with GARCH or SV, in which the conditional variances of different shocks are modelled independently.

Another extension grants each Markov process the sparsity property. In this specification, the number of regimes is set higher than the hypothesised number of regimes in the data-generating process, and some of these regimes are allowed to have zero occurrences throughout the sample. It is inspired by the overfitting mixture model proposed by \cite{malsiner-walli2016} and adapted to Markov-switching models. Therefore, in our model, the number of regimes with non-zero occurrences is determined during estimation, as only these regimes contribute to the likelihood function. This comes at a cost of the regime-specific conditional variance parameters losing their interpretability. Instead, the sequence of their values over time has standard interpretations as in the GARCH and SV models. Our original extensions of the Markov-switching heteroskedasticity models make them more flexible and closer to these two alternatives, in which the shock-specific variances are allowed to change frequently as the data require. Importantly, Markov-switching models are much faster to estimate than SV. 

We facilitate such modelling by several numerical and inferential developments. Firstly, we complement the normalisation restriction proposed by \cite{brunnermeier2021a}, which we adopt in our model, by the derivation of a new distribution for normalised conditional variances which we call the Inverse Gamma-based Dirichlet distribution. This distribution is easy to sample from, facilitating Gibbs sampling, an estimation procedure much more efficient than the Metropolis-Hastings step used by \cite{brunnermeier2021a}. Secondly, a probability density function for this distribution is derived, facilitating fast computation of Bayes factors for homoskedasticity verification using Savage-Dickey density ratio, following the approach proposed by \cite{lutkepohl2020} and \cite{lutkepohl2025}. The latter study provides suitable conditions for partial identification of a structural shock through heteroskedasticity in the context of our model. Consequently, we use the new computation of Bayes factors for homoskedasticity to make probabilistic statements about whether a shock, the corresponding structural parameters, and impulse responses are identified through heteroskedasticity. Finally, we follow recent advances and make our model identified through heteroskedasticity order-invariant as in \cite{brunnermeier2021a}, \cite{chan2024}, and \cite{lutkepohl2025}. All of these methods are implemented in the \textbf{R} package \textbf{bsvars} by \cite{wozniak2025,wozniak2024}.

Our methodological developments lead to substantial improvements in estimation efficiency, reliability of homoskedasticity verification, and forecasting performance. Our Monte Carlo simulations document several important results. Firstly, our SVAR model with sparse HMSH estimates the structural matrix parameters more precisely than alternative models. This conclusion is challenged only by the SV model in large samples generated from the SVAR-SV data-generating process and by the stationary HMSH in small samples generated from the stationary HMSH process. Secondly, the sparse HMSH model normalises conditional variances most efficiently, a conclusion challenged only by SV models in large samples. Thirdly, the sparsity of our Markov-switching models is essential for precise verification of homoskedasticity. Importantly, our verification outperforms the procedure by \cite{lutkepohl2025} in terms of reliability across many data-generating processes.

Finally, we apply our family of models to a 10-variable system of macro-financial aggregates used by \cite{brunnermeier2021a}, updating it to the most recent observations, to document excellent forecasting performance and accurate structural analyses. We show that models featuring sparse Markov-switching volatility appear essential for precise point and density forecasting. Their forecasting performance is even better than that of SVARs with SV thus far considered as the top forecasting specification. We further show that the monetary policy shock is identified through heteroskedasticity and that its conditional standard deviation estimates, comparable to those reported by \cite{sims2006}, and its impulse responses  greatly resemble those reported by \cite{brunnermeier2021a} in the short- and mid-term.

This paper is dedicated to the memory of Professor Christopher Sims, who passed away on March 14, 2026. His contributions greatly influenced this work and the authors’ research agenda.

%
%

\section{The SVAR-HMSH Model}\label{sec:model}

\noindent This section scrutinises the SVAR model with Heterogeneous Markov-Switching Heteroskedasticity (SVAR-HMSH). We first introduce the SVAR model, followed by the specification of the error term distribution and the volatility model. Then, we discuss homoskedasticity verification, normalisation, and estimation.

\subsection{Structural Vector Autoregression}

\noindent Our SVAR model follows closely the specification proposed by \cite{lutkepohl2025}, including the hierarchical prior distributions that are are detailed in \ref{sec:priors}. The reduced-form equation is the VAR equation with $p$ lags specified for an $N$-vector $\mathbf{y}_t$ collecting observations on $N$ variables at time $t$:
\begin{align}
\mathbf{y}_t &= \mathbf{A}_1 \mathbf{y}_{t-1} + \dots + \mathbf{A}_p \mathbf{y}_{t-p} + \mathbf{A}_d \mathbf{d}_t +  \boldsymbol{\varepsilon}_t, \label{eq:var}
\end{align}
where $\mathbf{A}_i$ are $N\times N$ autoregressive matrices, $\mathbf{d}_t$ is a $D$-vector of exogenous variables that might include deterministic terms such as the constant term, $\mathbf{A}_d$~is an $N\times D$ matrix of the corresponding parameters, and $\boldsymbol{\varepsilon}_t$ collects the $N$ reduced-form error terms. Collect all the autoregressive matrices and the slope terms in an $N\times (Np+D)$ matrix $\mathbf{A} = \begin{bmatrix}\mathbf{A}_1& \dots & \mathbf{A}_p & \mathbf{A}_d\end{bmatrix}$ and the explanatory variables in a $(Np+D)$-vector $\mathbf{x}_t = \begin{bmatrix}\mathbf{y}_{t-1}' & \dots & \mathbf{y}_{t-p}' & \mathbf{d}_{t}' \end{bmatrix}'$. Then equation \eqref{eq:var} can be written in the matrix form as
\begin{align}
\mathbf{y}_t &= \mathbf{A}\mathbf{x}_t + \boldsymbol{\varepsilon}_t. \label{eq:rf}
\end{align}

The structural form equation determines the linear relationship between the reduced-form innovations $\boldsymbol{\varepsilon}_t$ and the structural shocks $\mathbf{u}_t$ using the $N\times N$ structural matrix $\mathbf{B}_0$:
\begin{align}
\mathbf{B}_0\boldsymbol{\varepsilon}_t = \mathbf{u}_t.\label{eq:sf}
\end{align}
The structural matrix specifies the contemporaneous relationships among the system's variables and determines the identification of the structural shocks in vector $\mathbf{u}_t$. Its appropriate construction may yield a specific interpretation of one or more of the shocks.

Finally, our model features time-varying volatility driven by HMSH shocks, facilitating identification. These characteristics are implemented by assuming that the structural shocks are jointly conditionally normal with zero mean and a diagonal covariance matrix given the parameters and the shock-specific Markov process, $s_{n.t}$. Therefore, the distribution of the $n\textsuperscript{th}$ structural shock at time $t$ is given by:
\begin{align}
u_{n.t}\mid s_{n.t}, \sigma_{n.s_{n.t}}^2 \sim\mathcal{N}\left( 0, \sigma_{n.s_{n.t}}^2 \right),\label{eq:ss}
\end{align}
where $\sigma_{n.s_{n.t}}^2$ denotes the conditional variance. In what follows, we present our extensions and complement them by a more detailed characterisation of the volatility process in Sections~\ref{ssec:hmsh}--\ref{ssec:volatility}. These novel extensions make the model more flexible, leading to more efficient estimation, reliable homoskedasticity verification, and excellent forecasting performance.

\subsection{Heterogeneous Markov-Switching Heteroskedasticity}\label{ssec:hmsh}

\noindent The time-variation of the conditional variances is determined by the shock-specific discrete-valued Markov process $s_{n.t}$ with $M_n$ regimes \citep[see][]{hamilton1989}. The shock-specific conditional variances change their values independently of each other according to their Markov processes, taking values $s_{n.t} = m\in{1,\dots,M_n}$. The properties of these Markov processes are determined by their corresponding transition matrix $\mathbf{P}_n$ which we discuss in Section~\ref{ssec:volatility}. An element of the transition matrix, $[\mathbf{P}_n]_{i.j}$, denotes the transition probability from regime $i$ to regime $j$ over the next period. The initial regime probabilities of this process are estimated and denoted by the $M_n$-vector $\boldsymbol{\pi}_{n.0}$.

Following \cite{brunnermeier2021a}, the regime-specific variances in the $n\textsuperscript{th}$ equation take value 1 on average:
\begin{align}
\frac{1}{M_n}\left(\sigma_{n.1}^2 + \dots + \sigma_{n.M_n}^2\right) = 1, \label{eq:normalisation}
\end{align}
which ensures the normalisation of the system. The normalisation is implemented by assuming a prior distribution for the variance parameters, leading to a posterior distribution that satisfies the restriction~\eqref{eq:normalisation} as well. Each of the variances has prior expected value 1, and their regimes have equal prior probabilities of occurrence equal to $M_n^{-1}$. Therefore, the prior for the conditional variances is the $M_n$-variate Dirichlet distribution:
\begin{align}
M_m^{-1}\left(\sigma_{n.1}^2, \dots, \sigma_{n.M_n}^2\right) \sim\mathcal{D}irichlet_M(\underline{e}_\sigma, \dots, \underline{e}_\sigma), \label{eq:sigmaMSprior}
\end{align}
where the hyper-parameter $\underline{e}_\sigma = 1$ is fixed.

The functional form of the full conditional posterior distribution of $\sigma_{n.s_{n.t}}^2$ is equivalent to that of an inverse gamma 2-distributed random variable. However, it does not incorporate the restriction \eqref{eq:normalisation}. To facilitate efficient Gibbs sampling that respects the restrictions and homoskedasticity verification, we define a new distribution.

\begin{df}[Inverse Gamma-based Dirichlet distribution]\label{df:ig2dirichlet}
Let $z_1,\dots,z_M$ be positive independent real-valued random variables each of which is distributed according to the inverse gamma 2 distribution \citep[see][]{bauwens1999}, $z_m\sim\mathcal{IG}2(s_m, \nu_m)$, for $m\in{1,\dots,M}$, where $s_m$ and $\nu_m$ are positive real numbers representing the scale and shape, respectively. The probability density function of $\mathcal{IG}2$ is given by:
\begin{equation*}
f_{\mathcal{IG}2}(z_m|s_m,\nu_m) \sim \mathcal{IG}2(s_m, \nu_m) = \Gamma\left( \frac{\nu_m}{2} \right)^{-1}\left( \frac{s_m}{2} \right)^{\frac{\nu_m}{2}}z_m^{-\frac{\nu_m+2}{2}}\exp\left(-\frac{1}{2}\frac{s_m}{z_m}\right),
\end{equation*}
where $\Gamma(\cdot)$ denotes the gamma function. Then, random variables $x_m = z_m/(z_1+\dots+z_M)$, for $m\in{1,\dots,M}$, follow the \emph{Inverse Gamma-based Dirichlet distribution} with probability density function given by:
\begin{equation*}
f_{\mathcal{IGD}}(x_1,\dots,x_M|s_1,\dots,s_M, \nu_1,\dots,\nu_M) = \frac{\Gamma\left( \frac{\sum_{m=1}^{M}\nu_m}{2} \right)}{\prod_{m=1}^{M}\Gamma\left( \frac{\nu_m}{2} \right)}\left( \prod_{m=1}^{M}s_m^{-1} \right) \left[ \prod_{m=1}^{M}\left( \frac{s_m}{x_m} \right)^{\frac{\nu_m+2}{2}} \right] \left( \sum_{m=1}^{M}\frac{s_m}{x_m} \right)^{-\frac{\sum_{m=1}^{M}\nu_m}{2}}.
\end{equation*}
\end{df}
\begin{algo}[Random number generator]\label{algo:ig2dirichlet}
To generate a random draw from the $M$-variate Inverse Gamma-based Dirichlet distribution $f_{\mathcal{IGD}}(x_1,\dots,x_M|s_1,\dots,s_M, \nu_1,\dots,\nu_M)$:
\begin{enumerate}
\item draw $z_m\sim\mathcal{IG}2(s_m,\nu_m)$, for $m\in{1,\dots,M}$, and
\item return $(x_1, \dots, x_M)$, where $x_m = z_m/(z_1+\dots+z_M)$.
\end{enumerate}
\end{algo}
\begin{proof}
The proofs of both the probability density function and the random number generator are based on the change of variables. It requires stating the relationships: $x = \sum_{m=1}^{M}z_n$  and $z_1= x_1x, \dots, z_{M-1} = x_{M-1}x, z_n = (1-x_1-\dots-x_{M-1})x = x_nx$ with the Jacobian of the transformation being equal to $x^{M-1}$. In the final step, $x$ needs to be integrated out of the constructed function, yielding our new density function.
\end{proof}
In the same way as $M$-dimensional Dirichlet-distributed random variables are constructed by normalising $M$ independent gamma-distributed ones when the degrees of freedom parameters are allowed to vary for each of these variables, but their scale parameters are all set to 1, the Inverse Gamma-based Dirichlet-distributed random variables are obtained by normalising inverse gamma 2-distributed ones that have unrestricted scale and shape parameters. Moreover, our new distribution belongs to a class of Generalised Liouville distributions \citep[see][]{rayens1994} whose general definition excludes applications as required in our work.

The full conditional posterior distribution for the shock-specific conditional variances with the normalising restriction \eqref{eq:normalisation} and following the prior distribution in \eqref{eq:sigmaMSprior} is given by the $M_n$-variate Inverse Gamma-based Dirichlet distribution. This result facilitates fast and efficient estimation of conditional variances via Gibbs sampling. This introduces a significant improvement over the estimation procedure used by \cite{brunnermeier2021a}, which involves a much less efficient Metropolis-Hastings step for the variance as well as for the structural matrix, $\mathbf{B}_0$.

\subsection{Homoskedasticity Verification Using Bayes Factor}\label{ssec:homo}

\noindent Our volatility model specification allows us to investigate identification of the $n^{\text{th}}$ shock through heteroskedasticity by verifying the hypothesis of homoskedasticity represented by a restriction 
\begin{align}
\sigma_{n.1}^2 = \dots = \sigma_{n.M_n}^2 = 1. \label{eq:hypothesishomo}
\end{align}
This is facilitated by combining several recent results and our novel developments. Firstly, \cite{lutkepohl2025} provide suitable general conditions for partial identification of a~structural shock through heteroskedasticity in their Theorem~A.1. These conditions state that a structural shock, the corresponding row of the $\mathbf{B}_0$ matrix, and impulse responses of this shock to all variables are identified through heteroskedasticity iff the sequence of conditional variances of this shock differs from those of other shocks, and are directly applicable in our setup. According to this conditon the only homoskedastic shock in the system is identified, and a heteroskedastic shock is identified with probability one. Note that the previously known conditions for the homogenous Markov-switching model proposed by \cite{lutkepohl2020} do not apply in our setup with heterogeneous Markov-switching.

We rely on a Bayes factor to verify the hypothesis \eqref{eq:hypothesishomo} and follow \cite{lutkepohl2020} and \cite{lutkepohl2025} to estimate it using the SDDR \citep[see][]{verdinelli1995} given by:
\begin{align}
SDDR_H = \frac{
p\left(\sigma_{n.1}^2 = \dots = \sigma_{n.M_n}^2 = 1 \mid \mathbf{y}\right)}{
p\left(\sigma_{n.1}^2 = \dots = \sigma_{n.M_n}^2 = 1\right)}. \label{eq:sddr}
\end{align}
The denominator of the ratio is the prior ordinate at the restriction \eqref{eq:hypothesishomo} and is straightforward to compute using the Dirichlet prior distribution in \eqref{eq:sigmaMSprior}. The numerator contains the marginal posterior distribution of the $n^{\text{th}}$ structural shock variances evaluated at the same restriction. It can be computed from the inverse gamma-based Dirichlet full conditional posterior distribution given in Definition \ref{df:ig2dirichlet} using the numerical integration technique proposed by \cite{gelfand1990} relying on the output from the Gibbs sampler. Consequently, the SDDR can be computed quickly given a sample from the posterior distribution.

The SDDR is interpreted as the Bayes factor for the restricted model relative to the unrestricted one. Equation \eqref{eq:sddr} shows that the posterior probability mass being more concentrated around the restriction than the prior distribution leads to the SDDR value greater than one and provides evidence in favour of the restrictions. In the opposite case, the ratio is less than one and evidence against the hypothesis.

\subsection{Alternative Volatility Specifications}\label{ssec:volatility}

\noindent Our flexible model specification for the volatility process nests several alternatives distinguished by different specifications of the Markov process.\footnote{Other possible alternatives, including those differing by the transition matrix specification, could be considered and were recently reviewed by \cite{song2021}.} We present them starting from our novel and most general model and then discuss the special cases. 

Our Heterogeneous Markov-Switching Heteroskedasticity model, abbreviated HMSH, is a novel proposal that generalises existing approaches in which SVARs were complemented with homogeneous Markov-switching heteroskedasticity \citep[see][]{sims2006,lanne2010,lutkepohl2020}, thereby making the model more flexible. We provide this model in two versions. The first assumes that the Markov process $s_{n.t}$ is stationary, irreducible and aperiodic for all $n$. These assumptions imply, amongst other properties, that all regimes must entail a strictly positive number of periods assigned to it over the sample in the estimation process. We follow \cite{fruhwirth-schnatter2006} and impose this restriction by requiring at least three observations to be assigned to each of the regimes at each iteration of the Gibbs sampler. This specification entertains a parametric interpretation in which the regimes can be empirically characterised by analysing the periods of high probability of occurrence for each regime and the regime-specific conditional variance estimates. In our estimations, we will use a model with $M_1 = \dots = M_N = 2$ regimes and refer to it by HMSH(2).

The second version of the HMSH model is a sparse representation that fixes the number of regimes at an overfitting value, say, $M_1 = \dots = M_N = 20$, and resigns from the Markov process' stationarity and irreducibility. Therefore, some regimes may have zero occurrences throughout the sample, allowing the number of regimes with non-zero occurrences to be estimated following the ideas of \cite{malsiner-walli2016} for a~mixture of normal distributions model. Due to its construction, the sparse HMSH model excludes regime-specific interpretation of parameters. Instead, the estimated sequence of conditional variances, $\sigma_{n.s_{n.t}}^2$, analysed over time $t=1,\dots,T$, enjoys standard interpretations. In our estimations, we refer to this model by HMSH(20).

A simplified model with homogeneous Markov-Switching Heteroskedasticity (MSH) is a special case of the HMSH model in which the Markov process is common across all shock variances, $s_t = s_{1.t} = \dots = s_{N.t}$. Therefore, in this model, the variances of all structural shocks change simultaneously, and the regime occurrence probabilities are estimated. This model is less flexible than the HMSH model, but it was nevertheless used in important studies by \cite{sims2006} and \cite{lanne2010}. We are using this model in its stationary and sparse representations, referred to as MSH(2) and MSH(20), respectively.

Finally, we also consider a model with exogenous heteroskedasticity (EXH), in which all conditional variances change at periods predetermined by the investigator. This amounts to fixing the Markov process $s_t$ before estimating the model. As long as it offers a seemingly too parsimonious representation for complicated patterns of macroeconomic series volatility, the model was used in the influential studies by
\cite{rigobon2003} and \cite{brunnermeier2021a}.

\subsection{Normalisation of the Structural Model}\label{ssec:normalise}

\noindent Normalisation of a structural heteroskedastic system is challenging because the structural matrix and conditional variances are embedded in a multiplicative relationship forming the predictive density covariance matrix given by:
\begin{align}
\boldsymbol\Sigma_t = \mathbf{B}_0^{-1}\diag\left(\boldsymbol\sigma_t^2\right)\mathbf{B}_0^{-1\prime},
\end{align}
where $\boldsymbol\sigma_t^2 = \begin{bmatrix}\sigma_{1.s_{1.t}}^2 & \dots & \sigma_{N.s_{N.t}}^2\end{bmatrix}'$. Consequently, in the absence of further restrictions, the rows of $\mathbf{B}_0$, and the corresponding elements of $\boldsymbol\sigma_t^2$ are identified up to a constant that changes their scaling. In Bayesian inference, the values around which the posterior sampler will settle are determined by the interplay between their prior distributions, not by the data, because the scaling problem here is likelihood-invariant.

In our model, the elements of $\mathbf{B}_0$ are unrestricted. Therefore, conditional variances are subject to normalisation. This is achieved by the normalising restriction~\eqref{eq:normalisation} proposed by \cite{brunnermeier2021a} that centers the $n\textsuperscript{th}$ shock conditional variances around its unconditional variance given by
\begin{align}
\mathbb{E}\left[\boldsymbol\sigma_{n.s_{n.t}}^2\right] = \pi_{n.1}\sigma_{n.1}^2 + \dots + \pi_{n.M_n}\sigma_{n.M_n}^2,
\end{align}
where $(\pi_{n.1}, \dots, \pi_{n.M_n})$ are regime ergodic probabilities \citep[see e.g.][]{song2021}. This centring is different from a common practice of centring the unconditional variance at value 1 \citep[see e.g.][]{normandin2004} or close to it \citep[see e.g.][]{lutkepohl2025}, but it provides normalisation.

\subsection{Estimation}

\noindent The model presented in the current section, together with the prior distributions in \ref{sec:priors} lead to an efficient Gibbs sampling algorithm leveraging the recent developments in the Bayesian estimation of SVARs. The autoregressive parameters are sampled from a multivariate normal full conditional posterior distribution using the algorithm by \cite{carriero2022}. The structural parameters are sampled from a~generalised-normal distribution following \cite{waggoner2003}. We use our new distribution to sample the conditional variances. The initial probabilities of the Markov process and transition probabilities are sampled using a Dirichlet distribution as in \cite{fruhwirth-schnatter2006}, while the forward-smoothing-backward-sampling algorithm by \cite{chib1996} is employed to simulate the Markov processes. All estimation algorithms are implemented in the \textbf{R} package \textbf{bsvars} by \cite{wozniak2025,wozniak2024}, using \textbf{C++} code to facilitate convenient data analysis and fast computations.

\bigskip To summarise, our proposal for the volatility model is novel and extends existing approaches in several directions. Both extensions to heterogeneous and sparse Markov-switching heteroskedasticity are new and generalise the homogeneous stationary Markov processes used by \cite{sims2006}. We also improve estimation efficiency for the conditional variances by introducing the new Inverse Gamma-based Dirichlet distribution, which facilitates Gibbs sampling rather than a less efficient Metropolis-Hastings algorithm. The proper density function for the conditional variances facilitates valid verification of the homoskedasticity hypothesis using the SDDR. Finally, our modelling extensions disvalidate the identification conditions by \cite{lanne2010} and \cite{lutkepohl2020} proposed for homogeneous Markov-switching heteroskedasticity and require the use of the more general conditions by \cite{lutkepohl2025}. Overall, we propose a complete toolset for flexible Markov-switching heteroskedasticity modelling, offering advantages and opportunities for SVAR applications.

\section{Two Monte Carlo Experiments}\label{sec:mc}

\noindent To investigate the performance of our new methods with respect to estimation efficiency, normalisation, and homoskedasticity verification reliability under a potentially misspecified heteroskedasticity specification, we perform two Monte Carlo experiments. In these exercises, we generate synthetic data from various structural heteroskedastic data-generating processes (DGPs) and apply our methods to estimate the parameters and verify homoskedasticity. The DGPs include SVARs with four alternative volatility specifications. The estimated models include SVARs with HMSH and homogenous specifications, both in their sparse and stationary Markov-switching versions, as well as an exogenous heteroskedasticity model similar to that used by \cite{brunnermeier2021a}, and a model with SV by \cite{lutkepohl2025}.

Three lessons emerge from this exercise. Firstly, our SVAR model with sparse HMSH estimates the structural matrix parameters more precisely than alternative models. This conclusion is challenged only by the SV model in large samples generated from the SVAR-SV data-generating process and by the stationary HMSH in small samples generated from the stationary HMSH process. Secondly, the sparse HMSH model normalises conditional variances most efficiently, a conclusion challenged only by SV models in large samples. Thirdly, the sparsity of our Markov-switching models is essential for precise verification of homoskedasticity. Importantly, our verification outperforms the procedure by \cite{lutkepohl2025} using SV in reliability for several DGPs.

Our Monte Carlo exercise design closely follows that by \cite{lutkepohl2025}, and simulates 100 artificial bivariate datasets from various DGPs which vary by sample length $T$ and volatility process specification. We set the sample sizes to $T\in\{260,780\}$, corresponding to 65 years of quarterly and monthly data, respectively. To focus on the structural system, we use a simplified DGP that ignores the SVAR’s autoregressive coefficients and concentrates on the structural matrix $\mathbf{B}_0$ and the conditional variances $\boldsymbol\sigma^2_t$. Consequently, the DGP for this exercise is  given by:
\begin{align}
\mathbf{B}_0\mathbf{y}_t = \mathbf{u}_t,\quad \mathbf{u}_t\sim\mathcal{N}_2\left(\mathbf{0}_2,\diag\left(\boldsymbol\sigma_t^2\right)\right),\label{eq:dgpsstruc}
\end{align}
where $\mathbf{B}_0 = \begin{bmatrix}100 & 80 \\ -20 & 200\end{bmatrix}$, and $\mathbf{0}_2$ denotes a vector of two zeros. The four alternative processes for conditional variances $\boldsymbol\sigma_t^2$ are given by:
\begin{description}
\item[SV:] where $\sigma_{n.t}^2 = \exp\left(0.5{h}_{n.t}\right)$, $h_{n.t} = 0.92h_{n.t-1} + v_{n.t}$, $h_{n.0}=0$, and $v_{n.t}\sim\mathcal{N}(0,1)$

\item[GARCH:] where $\sigma_{n.t}^2 = 0.02 + 0.28u_{n.t-1}^2 + 0.7\sigma_{n.t-1}^2$ and $\sigma_{n.0}^2 = 1$.

\item[MSH(2):] where $\sigma_{n.t}^2 = \sigma_{n.s_t}^2$, $s_t$ is a two-state Markov process with transition probabilities $\mathbf{P}=\begin{bmatrix}0.98&0.02\\0.02&0.98\end{bmatrix}$, and variances $\sigma_{1.s_t = 1}^2 = 1.99$, $\sigma_{1.s_t = 2}^2 = 0.01$, $\sigma_{2.s_t = 1}^2 = 0.85$, and $\sigma_{2.s_t = 2}^2 = 1.15$.

\item[HMSH(2):] where $\sigma_{n.t}^2 = \sigma_{n.s_{n.t}}^2$, each $s_{n.t}$ is a two-state Markov process with equal transition probabilities $\mathbf{P}_n=\begin{bmatrix}0.98&0.02\\0.02&0.98\end{bmatrix}$, and variances $\sigma_{1.s_{1.t} = 1}^2 = 1.99$, $\sigma_{1.s_{1.t} = 2}^2 = 0.01$, $\sigma_{2.s_{2.t} = 1}^2 = 0.85$, and $\sigma_{2.s_{2.t} = 2}^2 = 1.15$.
\end{description}
Note that the variances in the two-regime Markov-switching DGPs sum to 2 for each shock in line with normalisation from equation~\eqref{eq:normalisation}.

For each of the artificially generated data sets we fit a structural model specified in expression~\eqref{eq:dgpsstruc} with the elements of the structural matrix following the zero-mean normal distribution with variance equal to 1000, and one of the six volatility processes: \textbf{EXH} -- an exogeneous regime change model with regime changes every 130 observations, \textbf{MSH(20)} -- a sparse MSH model with twenty regimes, \textbf{MSH(2)} -- a stationary MSH model with two regimes, \textbf{HMSH(20)} -- a sparse HMSH model with twenty regimes, \textbf{HMSH(2)} -- a~stationary HMSH model with two regimes, and \textbf{SV} with non-centred Stochastic Volatility by \cite{lutkepohl2025}.

\subsection{Estimation Efficiency Under Misspecified Heteroskedasticity}\label{sec:mceff}

\noindent We first investigate the efficiency of structural matrix estimation for models estimated using data generated by DGPs in which both shocks are heteroskedastic and follow a~specific volatility process. In this case, the parameters of matrix $\mathbf{B}_0$ are identified. Therefore, estimation efficiency is investigated by analysing the ratios of root-mean-squared estimation errors (RMSEs) for all elements of the estimated matrix. These ratios include the RMSE of a particular model relative to one of the HMSH(20) models and are reported in Table~\ref{tab:mceffB}. Therefore, their values greater than 1 indicate that the parameters in our HMSH(20) model are estimated more precisely than in other models.

\begin{table}[h!]
\begin{center}
\caption{Simulation results: relative RMSEs of the structural marix $\mathbf{B}_0$ for alternative volatility models}
\label{tab:mceffB}
\begin{tabular}{ll|cccccc}
  \toprule
 && \multicolumn{6}{c}{Estimated Models} \\
$T$ & DGPs & EXH & MSH(20) & MSH(2) & HMSH(20) & HMSH(2) & SV \\ [1ex]
  \midrule
\multirow{4}{*}{260}  & SV & 1.37 & 1.01 & 1.51 & 1.00 & 1.02 & 1.08 \\ 
  & GARCH & 1.26 & 0.97 & 1.46 & 1.00 & 1.12 & 1.47 \\ 
  & MSH(2) & 1.64 & 1.01 & 1.66 & 1.00 & 0.61 & 2.46 \\ 
  & HMSH(2) & 1.73 & 1.02 & 1.71 & 1.00 & 0.60 & 2.44 \\[1ex]
  \midrule
\multirow{4}{*}{780}    & SV & 1.04 & 1.01 & 1.36 & 1.00 & 1.02 & 0.81 \\ 
  & GARCH & 1.00 & 1.02 & 1.27 & 1.00 & 1.03 & 1.24 \\ 
  & MSH(2) & 1.25 & 1.06 & 2.48 & 1.00 & 1.34 & 2.02 \\ 
  & HMSH(2) & 1.44 & 1.06 & 2.46 & 1.00 & 1.35 & 2.10 \\[1ex] 
   \bottomrule
\end{tabular}
\end{center}

\smallskip{\footnotesize \textit{Note:} The table reports the ratios of the overall root-mean-squared error (RMSE) for all the elements of the structural matrix, $\mathbf{B}_0$. They are computed based on 100 artificial data sets generated for each of the four DGPs, namely SV, GARCH, MSH(2), and HMSH(2), and for two sample sizes, $T=260$ and $T=780$. The RMSEs are computed for the structural models with EXH, MSH(20), MSH(2), HMSH(20), HMSH(2) and SV volatility specifications. Reported RMSE ratios are computed relative to that of the model with HMSH(20) volatility specification. Therefore, their values greater than 1 indicate that the parameters in our sparse SVAR-HMSH are estimated more precisely than in other models.}
\end{table}

Our HMSH(20) model indeed features excellent estimation efficiency for the structural matrix parameters across all DGPs and sample sizes. Another two models nested within our general specification, namely, sparse MSH(20) and stationary heterogeneous HMSH(2), also perform well. Other specifications, including the stationary homogeneous MSH(2) model, EXH, and SV, tend to fail essentially for particular DGPs, such as the SV model for regime-change DGPs: MSH(2) and HMSH(2). Noticeably, the SV model performs excellently for SV DGP in large samples.

We further focus on whether the models are efficient in normalising the structural system. Therefore, we report ratios of RMSEs for the sequences of conditional variances across the shocks and over time, $\boldsymbol\sigma_t$, for all $t=1,\dots,T$ in Table~\ref{tab:mceffsigma}.

\begin{table}[h!]
\begin{center}
\caption{Simulation results: relative RMSEs of conditional standard deviation $\boldsymbol\sigma_t$ for all $t$ for alternative volatility models}
\label{tab:mceffsigma}
\begin{tabular}{ll|cccccc}
  \toprule
 && \multicolumn{6}{c}{Estimated Models} \\
$T$ & DGPs & EXH & MSH(20) & MSH(2) & HMSH(20) & HMSH(2) & SV \\ [1ex]
  \midrule
\multirow{4}{*}{260} & SV & 1.06 & 1.00 & 1.05 & 1.00 & 1.03 & 0.98 \\ 
  & GARCH & 1.07 & 1.00 & 1.05 & 1.00 & 1.06 & 1.04 \\ 
  & MSH(2)  & 1.21 & 1.01 & 1.06 & 1.00 & 0.97 & 1.28 \\ 
  & HMSH(2) & 1.21 & 1.01 & 1.04 & 1.00 & 0.97 & 1.28 \\ [1ex] 
  \midrule
\multirow{4}{*}{780} & SV & 1.05 & 1.00 & 1.06 & 1.00 & 1.04 & 0.94 \\ 
  & GARCH & 1.02 & 1.00 & 1.05 & 1.00 & 1.04 & 0.89 \\ 
  & MSH(2) & 0.97 & 1.01 & 1.06 & 1.00 & 1.65 & 0.97 \\ 
  & HMSH(2) & 0.97 & 1.00 & 1.05 & 1.00 & 1.67 & 0.97  \\[1ex] 
   \bottomrule
\end{tabular}
\end{center}

\smallskip{\footnotesize \textit{Note:} The table reports the ratios of the overall root-mean-squared error (RMSE) for conditional standard deviations, $\boldsymbol\sigma_t$, for $t=1,\dots,T$. They are computed based on 100 artificial data sets generated for each of the four DGPs, namely SV, GARCH, MSH(2), and HMSH(2), and for two sample sizes, $T=260$ and $T=780$. The RMSEs are computed for the structural models with EXH, MSH(20), MSH(2), HMSH(20), HMSH(2) and SV volatility specifications. Reported RMSE ratios are computed relative to that of the model with HMSH(20) volatility specification. Therefore, their values greater than 1 indicate that the our sparse SVAR-HMSH model normalises the variances more efficiently than in other models.}
\end{table}

\noindent All the proposed models are excellent at estimating conditional variances, with the sparse models MSH(20) and HMSH(20) performing robustly across DGPs and sample sizes. In large samples, the SV model beats all competitors, and the EXH model performs surprisingly well. This is not the case only for HMSH(2) for regime-change GDPs in large samples, which is a surprising reversal of this model's excellent performance in small samples.

\subsection{Assessing the Verification Procedure for Partial Identification}\label{sec:mcass}

\noindent As discussed in \cite{lutkepohl2025}, verifying whether a shock is partially identified boils down to testing for homoskedasticity. To explore how well the SDDR procedure detects homoskedastic shocks, our second MC exercise evaluates its performance under an alternative volatility process in the DGPs. For each volatility process, we generate data using four different scenarios: (1) both shocks are homoskedastic, (2) the first shock is homoskedastic while the second shock is heteroskedastic, (3) the first shock is heteroskedastic while the second shock is homoskedastic, (4) both shocks are heteroskedastic. For homoskedastic shocks, we set $\sigma^2_{n.t}=1$ for all $t$, while the heteroskedastic shocks are generated by the four different
volatility models. Akin to the first MC exercise, we consider two sample sizes, $T\in\{260,780\}$, and generate 100 simulated datasets for each of the four shock scenarios described above.

Tables \ref{tab:mcass} and \ref{tab:mcass2} report rejection rates for testing the homoskedasticity of the first shock. These rates are obtained using two strategies for constructing critical values, which we label $l$-value and $q$-value following \cite{BJ95} and \cite{Storey02}. In the $l$-value approach (Panel A), we apply a decision-theoretic rule and reject homoskedasticity if $SDDR_{H}<1$, where $SDDR_{H}$ is defined in equation~\eqref{eq:sddr}. In this case, rejection occurs when the posterior assigns more than 50\% probability to heteroskedasticity. In the $q$-value approach (Panel B), the critical value corresponds to the fifth percentile of the Bayes factors, $SDDR_{H}$, computed under the null in equation~\eqref{eq:hypothesishomo} in 100 repetitions of the experiment. As a result, the rejection rate in the first row of Panel B is fixed at 0.05.

As both tables report rejection rates for testing homoskedasticity of the first shock, the first two rows for each sample in Panel A should be close to 0. This is the case for the sparse representations of models MSH(20) and HMSH(20), EXH and SV models, but not for stationary Markov-switching models MSH(2) and HMSH(2). The latter models exhibit high rejection rates for homoskedasticity even when the first shock is homoskedastic, a consequence of the stationary assumption and its implementation, which requires a minimum of 3 periods to be classified into each regime during estimation. We investigated parameter estimates for these models and found that the second regime included very few observations, mostly exactly equal to 3. Apparently, these models captured outlying observations, leading to high conditional variance estimates in the second regime, which, in turn, led to high rejection rates for homoskedasticity. This is not the case when the model is given sufficient flexibility, as in sparse Markov-switching models.

Sparsity is an essential feature of the models for detecting heteroskedasticity as well. The rejection rates reported in the third and fourth rows of the panels for each sample should be high. This is robustly the case only for sparse models, MSH(20) and HMSH(20). Even the SV model by \cite{lutkepohl2025}, which features excellent performance for GARCH and SV DGPs, fails to detect heteroskedasticity in regime-change DGPs, MSH, and HMSH.\footnote{\cite{lutkepohl2025} investigates their SDDR performance for Markov-switching DGPs. However, their conditional variances were not normalised in the DGPs. The discrepancy in our results and theirs can be attributed to this change in the parameter values of the DGPs only.} To summarise, our methodological developments make sparse Markov-switching models, including the sparse HMSH model, lead to reliable homoskedasticity verification and constitute a viable alternative to existing frontier methods.

\begin{landscape}

\begin{table}[h!]
\begin{center}
\caption{Simulation results: rejection rates for homoskedasticity using Bayes factor for EXH, MSH(20), and MSH(2) models}
\label{tab:mcass}
\small
\begin{tabular}{cp{2.5cm}|cccc|cccc|cccc}
  \toprule
 && \multicolumn{12}{c}{Estimated models}\\
  && \multicolumn{4}{c}{EXH}& \multicolumn{4}{c}{MSH(20)}& \multicolumn{4}{c}{MSH(2)}\\
\midrule
 && \multicolumn{4}{c}{DGPs}& \multicolumn{4}{c}{DGPs}& \multicolumn{4}{c}{DGPs}\\
$T$ & homoskedastic shocks in DGP & SV & GARCH & MSH & HMSH & SV & GARCH & MSH & HMSH & SV & GARCH & MSH & HMSH \\ [1ex]
  \midrule
\multicolumn{14}{c}{Panel A: $l$-value approach} \\[1ex]
\multirow{4}{*}{260} & 1 \& 2 & 0.01 & 0.01 & 0.01 & 0.01 & 0.06 & 0.06 & 0.06 & 0.06 & 0.81 & 0.81 & 0.81 & 0.81 \\ 
  & 1 & 0.09 & 0.09 & 0.04 & 0.04 & 0.04 & 0.03 & 0.05 & 0.05 & 0.05 & 0.20 & 0.78 & 0.78 \\[1ex] 
  & 2 & 0.60 & 0.49 & 0.63 & 0.63 & 0.99 & 0.81 & 0.96 & 0.96 & 0.91 & 0.91 & 0.98 & 0.98 \\
  & none & 0.67 & 0.56 & 0.61 & 0.60 & 1.00 & 0.81 & 0.95 & 0.98 & 0.99 & 0.83 & 0.95 & 0.98 \\[1ex] 
\midrule
\multirow{4}{*}{780} & 1 \& 2 & 0.00 & 0.00 & 0.00 & 0.00 & 0.14 & 0.14 & 0.14 & 0.14 & 0.95 & 0.95 & 0.95 & 0.95 \\
  & 1 & 0.00 & 0.02 & 0.00 & 0.00 & 0.09 & 0.12 & 0.12 & 0.12 & 0.01 & 0.05 & 0.94 & 0.94 \\ [1ex]
  & 2 & 0.99 & 0.98 & 0.21 & 0.21 & 1.00 & 0.99 & 0.57 & 0.57 & 0.98 & 0.99 & 0.98 & 0.98 \\ 
  & none & 0.99 & 0.98 & 0.21 & 0.19 & 1.00 & 0.99 & 0.49 & 0.55 & 1.00 & 0.95 & 0.98 & 0.98 \\ [1ex]
  \midrule
\multicolumn{14}{c}{Panel B: $q$-value approach} \\[1ex]
\multirow{4}{*}{260} & 1 \& 2 & 0.05 & 0.05 & 0.05 & 0.05 & 0.05 & 0.05 & 0.05 & 0.05 & 0.05 & 0.05 & 0.05 & 0.05 \\ 
  & 1 & 0.16 & 0.16 & 0.10 & 0.10 & 0.02 & 0.01 & 0.03 & 0.03 & 0.00 & 0.01 & 0.04 & 0.04 \\[1ex] 
  & 2 & 0.66 & 0.55 & 0.65 & 0.65 & 0.99 & 0.80 & 0.96 & 0.96 & 0.85 & 0.48 & 0.82 & 0.82 \\ 
  & none & 0.78 & 0.61 & 0.64 & 0.66 & 1.00 & 0.78 & 0.95 & 0.97 & 0.88 & 0.37 & 0.79 & 0.82 \\ [1ex]
\midrule
\multirow{4}{*}{780} & 1 \& 2 & 0.05 & 0.05 & 0.05 & 0.05 & 0.05 & 0.05 & 0.05 & 0.05 & 0.05 & 0.05 & 0.05 & 0.05 \\
  & 1 & 0.04 & 0.12 & 0.04 & 0.04 & 0.05 & 0.06 & 0.08 & 0.08 & 0.00 & 0.00 & 0.04 & 0.04 \\[1ex] 
  & 2 & 0.99 & 0.99 & 0.42 & 0.42 & 1.00 & 0.99 & 0.42 & 0.42 & 0.98 & 0.75 & 0.11 & 0.11 \\ 
  & none & 1.00 & 1.00 & 0.42 & 0.45 & 1.00 & 0.99 & 0.36 & 0.37 & 0.98 & 0.59 & 0.03 & 0.07 \\ [1ex]
   \bottomrule
\end{tabular}
\end{center}

{\footnotesize \textit{Note:} The table reports rejection rates for the hypothesis of homoskedasticity in the first shock stated in restriction \eqref{eq:hypothesishomo} using simulated data. The rates are calculated based on 100 realizations of DGPs each with the following characteristics: sample sizes: $T\in\{260,780\}$; volatility processes: SV, GARCH, MSH(2), and HMSH(2); homoskedastic shock arrangements: shocks 1 \& 2, shock 1, shock 2, none.}
\end{table}
\end{landscape}

\begin{landscape}
\begin{table}[h!]
\centering
\caption{Simulation results: rejection rates for homoskedasticity using Bayes factor for HMSH(20), HMSH(2), and SV models}
\label{tab:mcass2}
\small
\begin{tabular}{cp{2.5cm}|cccc|cccc|cccc}
  \toprule
 && \multicolumn{12}{c}{Estimated models}\\
  && \multicolumn{4}{c}{HMSH(20)}& \multicolumn{4}{c}{HMSH(2)}& \multicolumn{4}{c}{SV}\\
\midrule
 && \multicolumn{4}{c}{DGPs}& \multicolumn{4}{c}{DGPs}& \multicolumn{4}{c}{DGPs}\\
$T$ & homoskedastic shocks in DGP & SV & GARCH & MSH & HMSH & SV & GARCH & MSH & HMSH & SV & GARCH & MSH & HMSH \\ [1ex]
  \midrule
\multicolumn{14}{c}{Panel A: $l$-value approach} \\[1ex]
\multirow{4}{*}{260} & 1 \& 2 &0.05 & 0.05 & 0.05 & 0.05 & 1.00 & 1.00 & 1.00 & 1.00 & 0.00 & 0.00 & 0.00 & 0.00 \\ 
  & 1 & 0.05 & 0.04 & 0.09 & 0.09 & 1.00 & 0.98 & 0.99 & 0.99 & 0.03 & 0.02 & 0.00 & 0.00 \\ [1ex]
  & 2 & 0.99 & 0.81 & 0.97 & 0.97 & 1.00 & 0.91 & 1.00 & 1.00 & 0.53 & 0.28 & 0.30 & 0.30 \\ 
  & none & 1.00 & 0.79 & 0.96 & 0.96 & 1.00 & 0.93 & 1.00 & 1.00 & 0.73 & 0.30 & 0.29 & 0.29 \\ [1ex]
\midrule
\multirow{4}{*}{780} & 1 \& 2 & 0.19 & 0.19 & 0.19 & 0.19 & 1.00 & 1.00 & 1.00 & 1.00 & 0.00 & 0.00 & 0.00 & 0.00 \\ 
  & 1 & 0.17 & 0.17 & 0.19 & 0.19 & 1.00 & 1.00 & 1.00 & 1.00 & 0.00 & 0.00 & 0.00 & 0.00 \\ [1ex]
  & 2 & 1.00 & 1.00 & 0.64 & 0.64 & 1.00 & 1.00 & 1.00 & 1.00 & 0.97 & 0.74 & 0.00 & 0.00 \\ 
  & none & 1.00 & 1.00 & 0.64 & 0.70 & 1.00 & 1.00 & 1.00 & 1.00 & 1.00 & 0.80 & 0.00 & 0.00 \\ [1ex]
  \midrule
\multicolumn{14}{c}{Panel B: $q$-value approach} \\[1ex]
\multirow{4}{*}{260} & 1 \& 2 & 0.05 & 0.05 & 0.05 & 0.05 & 0.05 & 0.05 & 0.05 & 0.05 & 0.05 & 0.05 & 0.05 & 0.05 \\ 
  & 1 & 0.05 & 0.04 & 0.09 & 0.09 & 0.04 & 0.02 & 0.06 & 0.06 & 0.10 & 0.15 & 0.08 & 0.08 \\ [1ex]
  & 2 & 0.99 & 0.80 & 0.96 & 0.96 & 0.00 & 0.04 & 0.93 & 0.93 & 0.81 & 0.51 & 0.53 & 0.53 \\ 
  & none & 1.00 & 0.79 & 0.96 & 0.96 & 0.00 & 0.02 & 0.89 & 0.90 & 0.96 & 0.63 & 0.50 & 0.52 \\ [1ex]
\midrule
\multirow{4}{*}{780} & 1 \& 2 & 0.05 & 0.05 & 0.05 & 0.05 & 0.05 & 0.05 & 0.05 & 0.05 & 0.05 & 0.05 & 0.05 & 0.05 \\ 
  & 1 & 0.03 & 0.05 & 0.04 & 0.04 & 0.02 & 0.01 & 0.03 & 0.03 & 0.12 & 0.15 & 0.08 & 0.08 \\ [1ex]
  & 2 & 1.00 & 0.99 & 0.37 & 0.37 & 0.01 & 0.05 & 0.08 & 0.08 & 0.97 & 0.97 & 0.19 & 0.19 \\ 
  & none & 1.00 & 0.99 & 0.24 & 0.37 & 0.05 & 0.05 & 0.05 & 0.07 & 1.00 & 0.98 & 0.19 & 0.18 \\ [1ex]
   \bottomrule
\end{tabular}

\smallskip{\footnotesize \textit{Note:} The note to Table~\ref{tab:mcass} applies.}
\end{table}

\end{landscape}

\section{Forecasting Study}\label{sec:forecast}

\noindent In this section, we compare the forecasting performance of SVARs with different volatility models in a recursive forecasting exercise for a 10-variable macro-financial system. Three key lessons emerge from it. First, some members of our sparse model family exhibit excellent density and point-forecast performance. Second, these sparse Markov-switching models outperform SV models in both density and point forecasts. Third, the evidence suggests that sparsity is the key driver of performance improvements, while heterogeneity is not prominent in this dataset.

We rely on the 10-variable monthly U.S.\ macro-financial system used by \cite{brunnermeier2021a}. This dataset includes industrial production $ip$, the PCE price index $p$, the federal funds rate $R$, household credit $hhc$, business credit $bc$, the Moody's Baa-Aaa corporate bond spread $GZ$, the spliced commercial paper rate spread $ES$, the term spread $TS$, the M2 monetary aggregate $m$, and Bloomberg's Commodity Index $pcm$. Industrial production, prices, credit, money, and the commodity index enter our data as 12-month log differences, while interest rates and spreads are in decimal form. The sample runs from January 1974 to September 2025. Relative to \cite{brunnermeier2021a}, whose sample extends through June 2015, our dataset differs in three ways: it spans an extended sample period, includes reconstructed series for money supply, commodity price index, and two financial stress indicators using alternative measurements due to the lack of original series availability, and uses 12-month log-differences where \cite{brunnermeier2021a} uses log-levels. We verified that our alternative measurements closely match the original dataset in the overlapping sample period.

We implemented a recursive expanding-window forecasting study, setting the first forecast origin to January 2016 and expanding the window by 1 month in each iteration. Our forecast performance measures include the mean predictive log score by \cite{ga10} for density forecasts and two measures, namely root-mean-squared forecast error (RMSFE) and mean-absolute forecast error (MAFE), for point forecasts. 

The competing specifications include exogenous heteroskedasticity (EXH) with pre-specified variance regimes, as in \cite{brunnermeier2021a}, with an extra regime during the six COVID months starting in March 2020; homogeneous Markov-Switching heteroskedasticity (MSH); and heterogeneous Markov-Switching heteroskedasticity (HMSH). As in the previous section, we consider  stationary and sparse representations for MSH and HMSH models with 2 and 20 regimes, respectively. We also include non-centred Stochastic Volatility by \cite{lutkepohl2025} as a benchmark referred to as SV. Our models share many features with those used by \cite{brunnermeier2021a}, such as an autoregressive lag order of 10, the estimation of all elements of the structural matrix, and some features of the Minnesota prior. However, our models differ by the specification of the prior mean for the structural matrix, set to the identity matrix multiplied by 100 in \cite{brunnermeier2021a} and set to a matrix of zeros consistent with the generalised normal distribution used by \cite{waggoner2003} and \cite{lutkepohl2025} in our models, hierarchical prior specification, and conditionally normal structural shocks, as opposed to Student-t shocks in \cite{brunnermeier2021a}.

\begin{table}[t!]
\begin{center}
\caption{Forecasting performance results: one-period-ahead density and point forecast measures for SVARs with various volatility specifications relative to that of SVAR with EXH.}
\label{tab:forecast_metrics}
\begin{tabular}{lrrr}
\toprule
Model & Difference of LPSs & Ratio of RMSFEs & Ratio of MAFEs \\
\midrule
HMSH(20)  & 0.179 & 0.983 & 0.971 \\
MSH(20)  & 1.595 & 0.986 & 0.973 \\
HMSH(2)  & -4.775 & 0.978 & 0.970 \\
MSH(2)  & 0.566 & 0.977 & 0.971 \\
SV &  0.761 & 0.988 & 0.980 \\
\bottomrule
\end{tabular}
\end{center}

\smallskip{\footnotesize \textit{Note:} The table reports forecasting performance measures for a 1-month-ahead forecast from a recursive expanding-window exercise. The difference of LPSs is the average log predictive score across all periods minus that computed for the SVAR with EXH. Its positive values mean that a model performs better than the benchmark EXH model. Ratios of RMSFEs and MAFEs are the point forecast measures relative to those of the benchmark model. Values less than 1 indicate improved performance relative to the benchmark.}

\end{table}

The forecasting performance measures reported in Table \ref{tab:forecast_metrics} show that, especially in terms of density forecasts, the sparse models outcompeted their counterpart stationary Markov-switching models. Here, sparsity plays a more decisive role than heterogeneity in improving forecasting performance. Selected sparse specifications within our family of models, the MSH(20) and HMSH(20) models, are competitive with, and in some cases better than, the SV model. In particular, MSH(20) improves LPS by 1.595 relative to EXH and outperforms the SV model in both density and point forecasts. The somehow relatively poor density forecasting performance of the HMSH(2) model boils down to worse performance on three dates, namely, March and April 2020 and March 2021. Excluding these dates from LPS computation levels this model's performance with other Markov-switching models.

At the one-step-ahead horizon, point forecasts are naturally very similar across models because they are based on the same information set and broadly similar conditional mean dynamics, but all our Markov-switching specifications outperform both SV and EXH in the point-forecast measures. The ratios relative to EXH also show non-negligible gains in the point forecasts precision across the Markov-switching specifications: the MAFE improves by about 3.0\% and RMSFE by about 2\%. Importantly, all our Markov-switching specifications outperform both SV and EXH on the point forecast measures.

\section{Monetary Policy Shock Identification}\label{sec:mps}

\noindent We continue the empirical investigation of SVARs identified via heteroskedasticity using the updated 10-variable system from Section~\ref{sec:forecast}. In what follows, we focus on the monetary policy shock and compare the findings from the full-sample estimation regarding identification, heteroskedasticity, and impulse responses across different volatility specifications. Following \cite{brunnermeier2021a}, the monetary policy shock is selected from the ten shocks identified through heteroskedasticity as the one that contributes most to the forecast error variance decomposition of the federal funds rate on impact. Subsequently, the posterior sample is normalised as in \cite{lutkepohl2025}. 

\begin{table}[t!]
\begin{center}
\caption{Homoskedasticity verification results: logarithms of Bayes factors for homoskedasticity of the monetary policy shock for SVARs with various volatility specifications.}
\label{tab:sddr}
\begin{tabular}{lr}
\toprule
Model & $\ln SDDR_H$ \\
\midrule
HMSH(20) & -240.283 \\
MSH(20) & -118.314 \\
HMSH(2) & -235.684 \\
MSH(2) & -1.245 \\
EXH & -54.291 \\
SV & -542.583 \\
\bottomrule
\end{tabular}
\end{center}

\smallskip{\footnotesize \textit{Note:} The table reports the logarithm of Bayes factors for the hypothesis of homoskedasticity of the monetary policy shock against the hypothesis of its heteroskedasticity. Negative values provide evidence against homoskedasticity. The Bayes factors for the first five models are computed using the SDDR from equation~\eqref{eq:sddr}, while that for the SV model is computed as in \cite{lutkepohl2025}.}

\end{table}

First, we apply the newly proposed methods to verify heteroskedasticity in the monetary policy shocks across SVARs, using the same six alternative volatility models as in Section~\ref{sec:forecast}. Table~\ref{tab:sddr} reports natural logarithms of Bayes factors for the hypothesis of homoskedasticity computed using the SDDR introduced in Section~\ref{ssec:homo}. The Bayes factors decisively reject homoskedasticity of the monetary policy shock in all models except for the SVAR with MSH(2). The more flexible the model used to verify homoskedasticity, the stronger the rejection of the hypothesis. The Bayes factors are the smallest for SV and heterogeneous Markov-switching models, HMSH(20) and HMSH(2). This is followed by a sparse, homogeneous MS model, MSH(20), and an exogenous regime-change model, EXH. For these models, evidence against homoskedasticity is \emph{very strong} according to the classification provided by \cite{kass1995}. For the MSH(2) model, the natural logarithm of the Bayes factor is equal to -1.25, which hardly exceeds \citeauthor{kass1995}'s threshold for \emph{positive} evidence. We label it as weak at best.

\begin{figure}[h!]
\begin{center}
\caption{Estimation results: time-varying conditional standard deviation of the monetary policy shock for SVARs with vaious volatility models.}

\smallskip
\includegraphics[scale = 0.7]{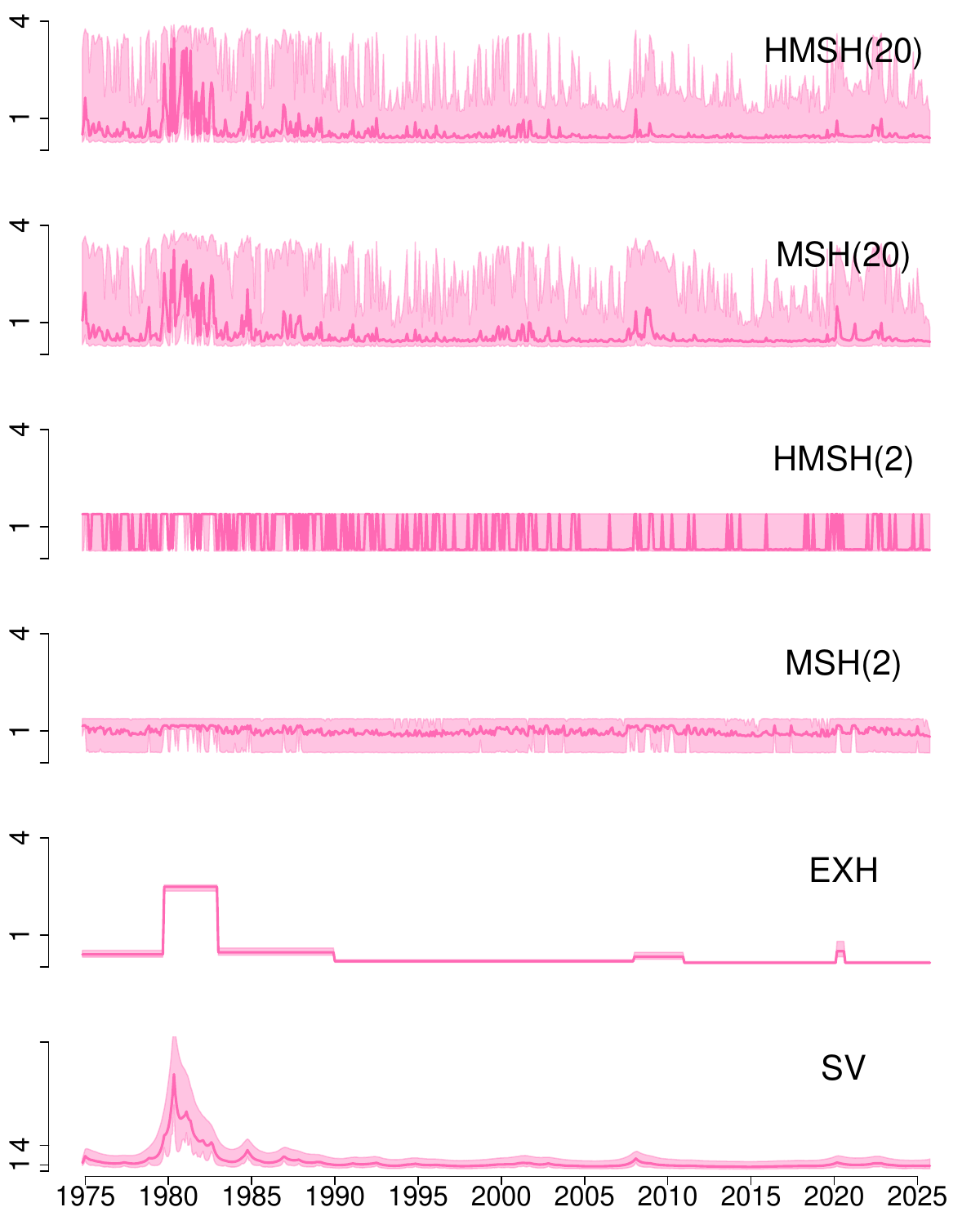}
\label{fig:csds}
\end{center}\footnotesize

Note: The figure reports the posterior mean with a bold line and 90\% highest posterior density intervals with the shaded area of the conditional standard deviation of the monetary policy shock for SVARs with six different volatility models.
\end{figure}

To provide further insights into the heteroskedasticity of the monetary policy shock, we plot its conditional standard deviations in Figure~\ref{fig:csds}. The figure shows similar volatility patterns estimated for most models with a very high volatility period in the early 1980s attributed by \cite{sims2006} to Paul Volcker's chairmanship at the Fed, and other much less pronounced volatility peaks following the dot-com bubble in the early 2000s, the global financial crisis in 2008, and the COVID pandemic beginning in March 2020. Models with high volatility process flexibility, such as SV and sparse MS models HMSH(20) and MSH(20), provide very similar estimates. The least flexible model, namely EXH, also features similar volatility patterns. However, this model benefits from the undeniable competence of the authors of \cite{brunnermeier2021a} in this topic. Another model, HMSH(20), produces somewhat constrained estimates that nevertheless follow similar patterns, with volatility estimates significantly different from 1 in quite some periods. Only the model deprived of sparsity or heterogeneity, namely MSH(2), fails to capture the heteroskedasticity of the monetary policy shock. This confirms our conclusion from Monte Carlo experiments in Section~\ref{sec:mc} that, given sufficient flexibility in the volatility model, our procedure can reliably verify the heteroskedasticity of the shock, whereas insufficient flexibility can lead to failure in this verification. Therefore, we strongly confirm that the US monetary policy shock is identified through heteroskedasticity, a~statement previously investigated by \cite{lutkepohl2020} using a 6-variable quarterly system.

\begin{figure}[t!]
\begin{center}
\caption{Estimation results: impulse responses of the first five variables to the monetary policy shock identified through heteroskedasticity using SVARs with various volatility specifications.}

\smallskip
\includegraphics[scale = 0.7]{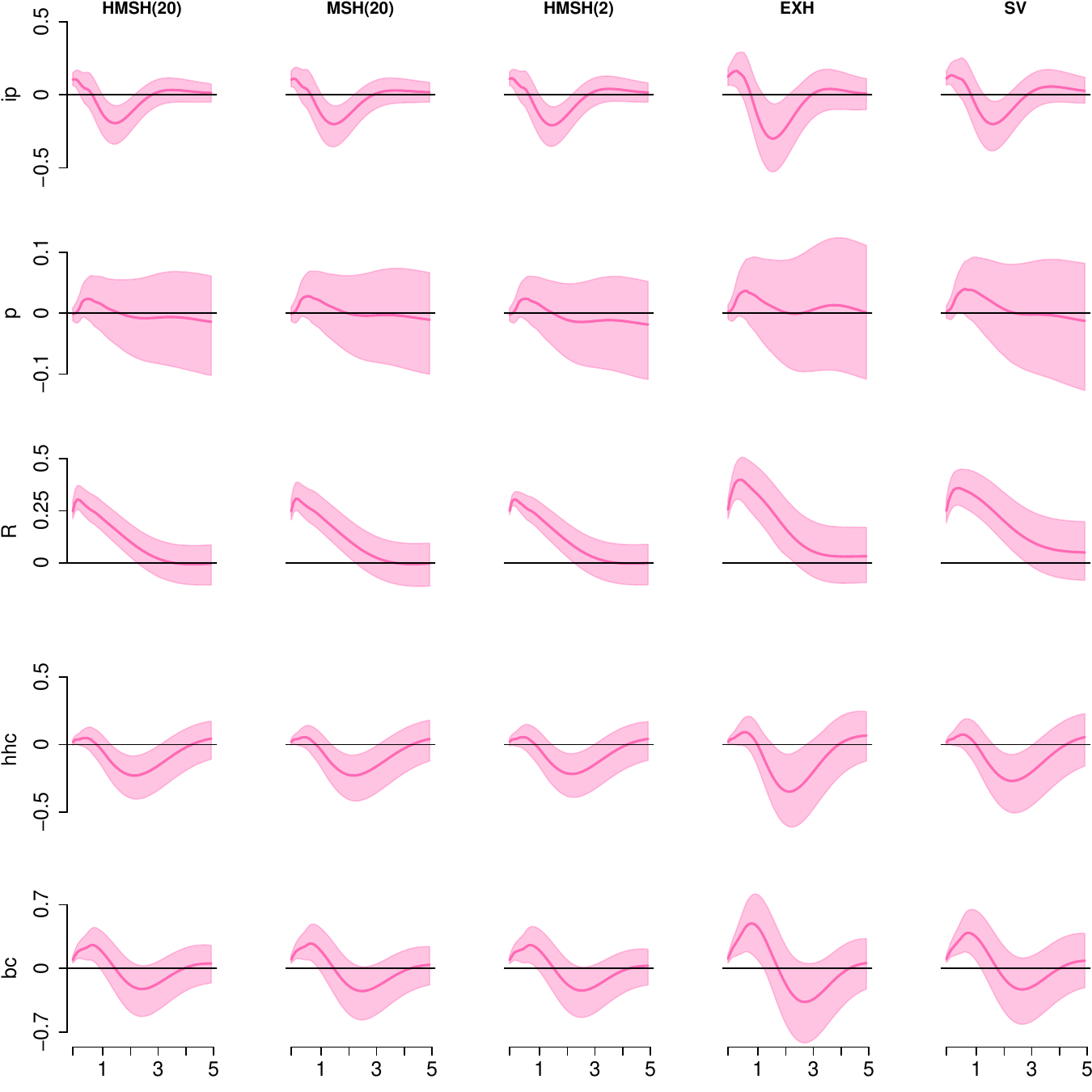}
\label{fig:irfs1}
\end{center}\footnotesize

Note: The figure reports the posterior mean with a bold line and 90\% highest posterior density intervals with the shaded area of the impulse responses of the monetary policy shock for 5-year horizon. Each row reports impulse responses of the same variable to the monetary policy shock. Each column reports impulse responses from the same SVAR model differenciated from others by its volatility specification. 
\end{figure}

\begin{figure}[t!]
\begin{center}
\caption{Estimation results: impulse responses of the last five variables to the monetary policy shock identified through heteroskedasticity using SVARs with various volatility specifications.}

\smallskip
\includegraphics[scale = 0.7]{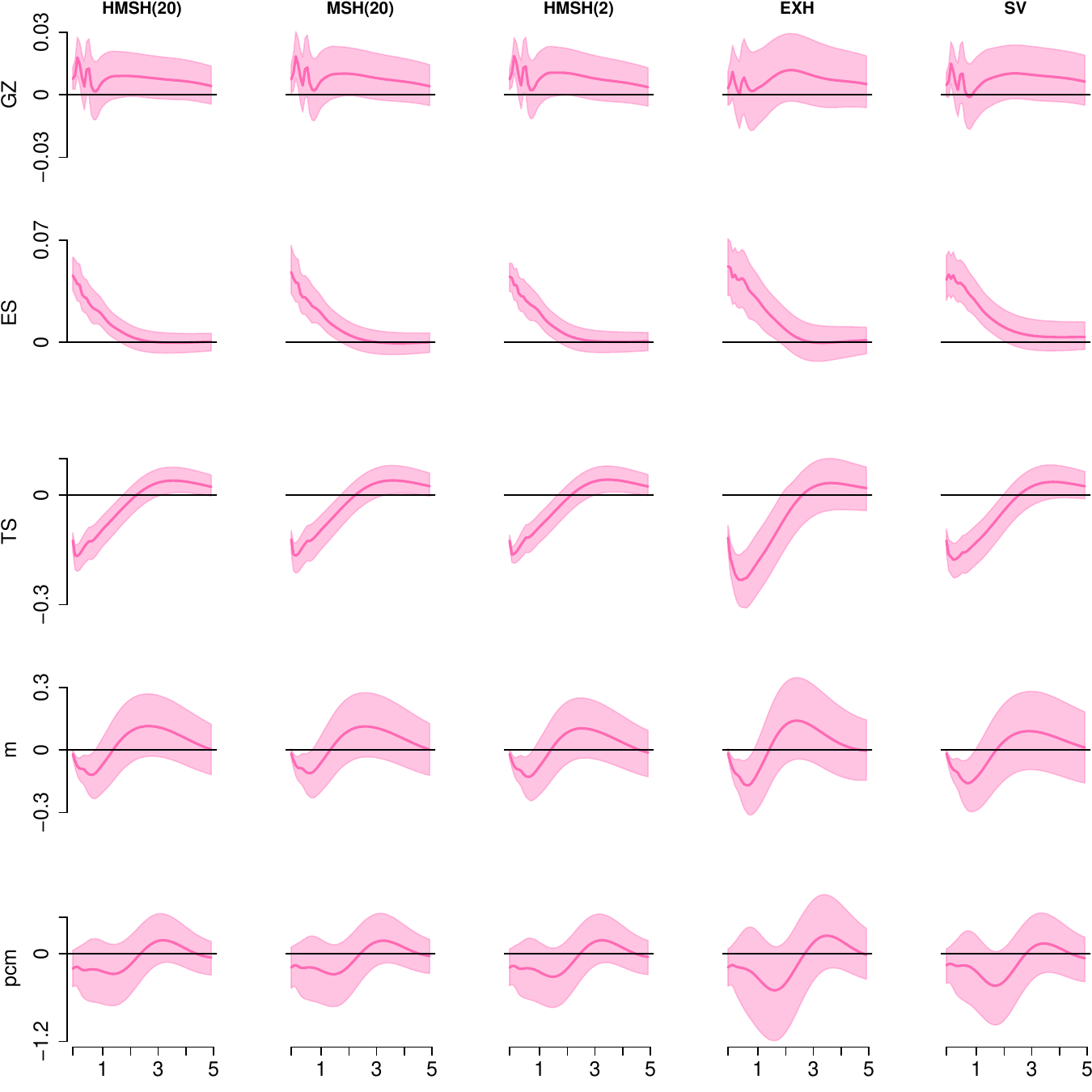}
\label{fig:irfs2}
\end{center}\footnotesize

Note: The note to Figure~\ref{fig:irfs1} applies. 
\end{figure}

Finally, we shift our focus to the effects of the monetary policy shock identified by time-varying volatility. Figures~\ref{fig:irfs1} and \ref{fig:irfs2} report the impulse responses of the first five and the last five variables in the system, respectively, to the monetary policy shock for a 5-year horizon. In these figures, we skip reporting the results for the MSH(2) model that fails to identify the monetary policy shock. Firstly, we point out the great level of similarity of the impulse responses across different models. Secondly, our impulse responses closely resemble those reported for the monetary policy shock by \cite{brunnermeier2021a} in the short- and mid-term, up to around 2,5 years. Beyond this horizon, our impulse responses become statistically insignificant, which we attribute to both the more flexible hierarchical prior we used for autoregressive parameters and differentiating some of the variables. Finally, we notice that the impulse responses for the least flexible volatility model, EXH, feature the widest highest posterior density intervals, which provides more nuance to the remark made by \cite{brunnermeier2021a}. With that respect, our results confirm that a model with a misspecified volatility process can successfully identify shocks through heteroskedasticity. However, this comes at a cost of estimation precision.

\section{Conclusions}\label{sec:conclusions}

\noindent We proposed a new class of sparse heterogeneous Markov-switching heteroskedasticity models suitable for SVARs and complement them with a range of numerical and inferential tools for fast and efficient estimation and reliable homoskedasticity verification. We show that these models are effective in normalising the structural system and estimating essential parameters. They also feature excellent forecasting performance, comparable to that of the Stochastic Volatility model, considered the best in the literature thus far. Finally, we apply the models to a 10-variable system of US macro-financial indicators and confirm that the US monetary policy shock is identified by heteroskedasticity, that the estimated impulse responses are similar across different volatility specifications, and that more flexible models provide more precise estimates.

\appendix

\section{Prior Distributions}\label{sec:priors}

\noindent Each of the rows of the matrix $\mathbf{A}$, denoted by $[\mathbf{A}]_{n\cdot}$, may feature exclusion restrictions representing, for instance, exogeneity, small-open economy assumptions, or no Granger causality hypothesis. The restrictions are imposed following the approach by \cite{waggoner2003} who decompose the row-specific parameters into:
\begin{align}
[\mathbf{A}]_{n\cdot} = \mathbf{a}_n\mathbf{V}_{A.n}, \label{eq:restrictionsA}
\end{align}
where $\mathbf{a}_n$ is a $1\times r_{A.n}$ vector collecting the elements to be estimated and the $r_{A.n}\times N$ matrix $\mathbf{V}_{A.n}$ including zeros and ones placing the estimated elements in the demanded elements of $[\mathbf{A}]_{n\cdot}$.

The unrestricted autoregressive parameters, $\mathbf{a}_n$, follow a multivariate conditional normal prior distribution, given the equation-specific shrinkage hyper-parameter $\gamma_{A.n}$, with the mean vector $\mathbf{V}_{A.n}\underline{\mathbf{m}}_{n.A}$ and the covariance $\gamma_{A.n}\mathbf{V}_{A.n}\underline{\Omega}_A\mathbf{V}_{A.n}'$, denoted by:
\begin{align}
\mathbf{a}_n'\mid\gamma_{A.n} \sim\mathcal{N}_{r_{A.n}}\left( \mathbf{V}_{A.n}\underline{\mathbf{m}}_{n.A}, \gamma_{A.n}\mathbf{V}_{A.n}\underline{\boldsymbol\Omega}_A\mathbf{V}_{A.n}' \right),\label{eq:priorA}
\end{align}
where $\underline{\mathbf{m}}_{n.A}$ is specified in-line with the Minnesota prior by \cite{doan1984} as a vector of zeros if all of the variables are stationary, or containing value 1 in its $n\textsuperscript{th}$ element if the $n\textsuperscript{th}$ variable is unit-root nonstationary. By default, $\underline{\boldsymbol{\Omega}}_A$ is a diagonal matrix with vector $\begin{bmatrix}\mathbf{p}^{-2\prime}\otimes\boldsymbol{\imath}_N' & 100\boldsymbol{\imath}_D'\end{bmatrix}'$ on the main diagonal, where $\mathbf{p}$ is a vector containing a sequence of integers from 1 to $p$ and $\boldsymbol\imath_N$ is an $N$-vector of ones. $\mathbf{V}_{A.n}$, $\underline{\mathbf{m}}_{n.A}$ and $\underline{\boldsymbol{\Omega}}_A$ can be modified by the user. This specification includes the shrinkage level exponentially decaying with the increasing lag order, relatively large prior variances for the deterministic term parameters, and the flexibility of the hierarchical prior that leads to the estimation of the level of shrinkage. The latter feature is facilitated by assuming a 3-level local-global hierarchical prior on the equation-specific reduced form parameters shrinkage given by 
\begin{align}
\gamma_{A.n} | s_{A.n}  &\sim\mathcal{IG}2\left(s_{A.n}, \underline{\nu}_A\right),\\
s_{A.n} | s_{A} &\sim\mathcal{G}\left(s_{A}, \underline{a}_A\right),\\
s_{A} &\sim\mathcal{IG}2\left(\underline{s}_{s_A}, \underline{\nu}_{s_A}\right),
\end{align}
where $\mathcal{G}$ and $\mathcal{IG}2$ are gamma and inverted gamma 2 distributions \citep[see][Appendix A]{bauwens1999}, hyper-parameters $\gamma_{A.n}$, $s_{A.n}$, and $s_{A}$ are estimated, and $\underline{\nu}_A$, $\underline{a}_A$, $\underline{s}_{s_A}$, and $\underline{\nu}_{s_A}$ are all set by default to value 10 to assure appropriate level of shrinkage towards the prior mean. The values of the hyper-parameters that are underlined in our notation can be modified by the user.

Zero restrictions can be imposed also on the structural matrix row-by-row following the framework proposed by \cite{waggoner2003} via the following decomposition of the $n\textsuperscript{th}$ row of the structural matrix, denoted by $[\mathbf{B}_0]_{n\cdot}$:
\begin{align}
[\mathbf{B}_0]_{n\cdot} = \mathbf{b}_n\mathbf{V}_{B.n}, \label{eq:restrictions}
\end{align}
where $\mathbf{b}_n$ is a $1\times r_{B.n}$ vector collecting the elements to be estimated and the $r_{B.n}\times N$ matrix $\mathbf{V}_{B.n}$ including zeros and ones placing the estimated elements in the demanded elements of $[\mathbf{B}_0]_{n\cdot}$.

The structural matrix $\mathbf{B}_0$ follows a conditional generalised-normal prior distribution by \cite{waggoner2003} that is proportional to:
\begin{align}
\mathbf{B}_0\mid\gamma_{B.1},\dots,\gamma_{B.N} \sim |\det(\mathbf{B}_0)|^{\underline{\nu}_B - N} \exp\left\{-\frac{1}{2} \sum_{n=1}^{N} \gamma_{B.n}^{-1} \mathbf{b}_n\underline{\boldsymbol\Omega}_{B.n}^{-1}\mathbf{b}_n' \right\},
\end{align}
where $\underline{\boldsymbol\Omega}_{B.n}$ is an $r_{B.n}\times r_{B.n}$ scale matrix set to the identity matrix by default, $\underline{\nu}_B \geq N$ is a shape parameter, and $\gamma_{B.n}$ is an equation-specific structural parameter shrinkage. The shape parameter $\underline{\nu}_B$ set to $N$ by default makes this prior a conditional, zero-mean $r_{B.n}$-variate normal prior distribution for $\mathbf{b}_n$ with the diagonal covariance and the diagonal element $\gamma_{B.n}$. The shape parameter can be modified by the user though.

This prior specification is complemented by a 3-level local-global hierarchical prior on the equation-specific structural parameters shrinkage given by 
\begin{align}
\gamma_{B.n} | s_{B.n}  &\sim\mathcal{IG}2\left(s_{B.n}, \underline{\nu}_b\right),\\
s_{B.n} | s_{B} &\sim\mathcal{G}\left(s_{B}, \underline{a}_B\right),\\
s_{B} &\sim\mathcal{IG}2\left(\underline{s}_{s_B}, \underline{\nu}_{s_B}\right),
\end{align}
where hyper-parameters $\gamma_{B.n}$, $s_{B.n}$, and $s_{B}$ are estimated and 
$\underline{\nu}_b$, $\underline{a}_B$, $\underline{s}_{s_B}$, and $\underline{\nu}_{s_B}$ are fixed to values 10, 10, 1, and 100 respectively to assure a flexible dispersed distribution \emph{a priori} but they can be modified by the user.

Finally, each of the rows of the transition matrix as well as the initial state probabilities follow the Dirichlet distribution:
\begin{align}
[\mathbf{P}_n]_{m\cdot} &\sim\mathcal{D}irichlet_{M_n}(\underline{e}, \dots, \underline{e})\\
\boldsymbol{\pi}_{n.0} &\sim\mathcal{D}irichlet_{M_n}(\underline{e}_0, \dots, \underline{e}_0).\label{eq:pi0prior}
\end{align}

\bibliographystyle{chicago}
\bibliography{bsvar_hmsh}

\end{document}